\documentstyle[12pt,aps]{revtex}
 \newcommand{\be}{\begin{equation}}
 \newcommand{\bea}{\begin{eqnarray}}
 \newcommand{\ee}{\end{equation}}
 \newcommand{\eea}{\end{eqnarray}}

\tightenlines
 \begin{document}
 \thispagestyle{empty}
 \begin{center}
 {\large\bf On the Schr\"odinger Equation for the Supersymmetric FRW model}
 \vspace{0.5cm} \\
 {\bf J.J. Rosales}\footnote{E-mail: juan@ifug3.ugto.mx},
 {\bf V.I. Tkach}\footnote{E-mail: vladimir@ifug3.ugto.mx}
 \vspace{0.5cm} \\
 {\it Instituto de F\'{\i}sica, Universidad de Guanajuato, \\
 Apartado Postal E-143, C.P. 37150, Le\'on, Gto. M\'exico}.
\vspace{0.5cm}\\
 {\bf A.I. Pashnev}\footnote{E-mail: pashnev@thsun1.jinr.ru} 
 \vspace{0.5cm}\\
{\it JINR-Bogoliubov Laboratory of Theoretical Physics\\
 141980 Dubna, Moscow Region, Russia}.
\vspace{0.5cm}\\

 \end{center}
 
 \vspace{0.5cm}

 \begin{abstract}
We consider a time-dependent Schr\"odinger equation for the
Friedmann-Robertson-Walker (FRW) model. We show that for this purpose it is
possible to include an additional action invariant under reparametrization
of time. The last one does not change the equations of motion for the 
minisuperspace model, but changes only the constraint. The same procedure is 
applied to the supersymmetric case.
  
\end{abstract}
\vspace{0.2cm}
PACS numbers: 04.65.+e, 98.80Hw
\newpage

\section{Introduction}
One of the most important questions in quantum cosmology is that of
identifying a suitable time parameter \cite{1} and a time-dependent
Wheeler-DeWitt equation \cite{2,3}. The main peculiarity of the
gravity theory is the presence of non-physical variables (gauge variables)
and constraints \cite{3,4,5,6}. They arise due to the general coordinate
invariance of the theory. The conventional Wheeler-DeWitt formulation
gives a time independent quantum theory \cite{7}. The canonical
quantization
of the minisuperspace approximation \cite{8} has been used to find results
in the hope, that they would illustrate the behaviour of general
relativity  \cite{9}. In the minisuperspace models \cite{2,7} there is a 
residual invariance under reparametrization of time (world-line symmetry). 
Due to this fact the equation that governs the quantum behaviour of these 
models is the Schr\"odinger equation for states with zero energy. On the 
other hand, supersymmetry transformations are more fundamental than time 
translations (reparametrization of time) in the sense, that these ones may be 
generated by anticommutators of the supersymmetry generators. The recent
introduction of supersymmetric mini\-superspace models has led to the square 
root equations for states with zero energy \cite{10,11,12}. The
structure of the world-line supersymmetry transformations has led to the 
zero Hamiltonian phenomena \cite{2,6,12}.
Investigations about time evolution problem for such quantum systems have
been carried in two directions: the cosmological models of gravity
have been quantized by reducing the phase space degrees of freedom
\cite{13,14,15,16,17} and with the help of the WKB approach
\cite{18,19,20}.

In this work we consider a time-dependent Schr\"odinger equation for the
homogeneous cosmological models. In our approach this equation arises due
to an additional action invariant under time reparametrization. The last
one does not change the equations of motion, but the constraint which
becomes time-dependent Schr\"odinger equation. In the case of the 
supersymmetric minisuperspace model we obtain the supersymmetric constraints, 
one of them is a square root of time-dependent Schr\"odinger equation.

The paper is organized as follows: In section 2, we applied the canonical
quantization procedure for the reparametrization invariant action. The
extension to supersymmetric FRW model is performed in section 3.

\section{Reparametrization Invariance}

We begin by considering an homogeneous and isotropic metric defined by
the line element

\be
d s^2 = - N^2 (t) dt^2 +R^2 (t) d\Omega^2_3 ,
\label{1}
\ee
 where the only dynamical degree of freedom is the scale factor $R(t)$.
 The lapse function $N(t)$, being a pure gauge variable, is not dynamical.
 The quantity $d \Omega^2_3$ is the standard line element on the unit
 three-sphere. We shall set $c= \hbar=1$. The pure gravitational action
 corresponding to the metric (\ref{1}) is
 \be
 S_g= \frac{6}{\kappa^2} \int \left( -\frac{R\dot R^2}{2N} +\frac{1}{2}
 k NR \right) dt ,
 \label{2}
 \ee
 where $k =1,0,-1$ corresponds to a closed, flat or open space.
 $\kappa^2 = 8\pi G_N $, where $G_N $ is the Newton's constant of
 gravity, and the overdot denotes differentiation with respect to $t$.
The action (\ref{2}) preserves the invariance under time reparametrization
 \be
 t^\prime \to t + a (t),
 \label{3}
 \ee
 in this case, $N(t)$ and $R(t)$ transform as
 \be
 \delta R = a\dot R \qquad\qquad \delta N= (aN)^{.},
 \label{4}
 \ee
 that is $R(t)$ transforms as a scalar and $N(t)$ as a one-dimensional
 vector, and its dimensionality is the inverse of $a(t)$.

 So, we consider the interacting action for the homogeneous real scalar
 matter field $\phi(t)$ and the scale factor $R(t)$. The action has the form
 \be
 S_m = \int \left( \frac{R^3 \dot\phi^2}{2N} - NR^3 V(\phi) \right) dt,
 \label{5}
 \ee
 this action is invariant under the local transformation (\ref{3}), if in
 addition to law transformations for $R(t)$ and $N(t)$ in (\ref{4}), the
matter field transforms as a scalar $\delta\phi = a \dot\phi$. Thus, our
 system is described by the full action
 \be
 S= S_g + S_m = \int \left( -\frac{3R{\dot R}^2}{\kappa^2 N} +
 \frac{R^3 {\dot \phi}^2}{2N} + \frac{3 k N R}{\kappa^2} -
 NR^3 V(\phi) \right)dt.
 \label{6}
 \ee
 Now, we shall consider the Hamiltonian analysis of this action. The
 canonical momenta for the variables $R$ and $\phi$ are given
 by
 \be
 P_R = \frac{\partial L}{\partial\dot R} = - \frac{6R \dot R}{\kappa^2 N},
 \qquad P_{\phi} = \frac{R^3 \dot \phi}{N}.
 \label{7}
 \ee
 Their canonical Poisson brackets are defined as
 \be
 \lbrace R, P_R \rbrace = 1, \qquad\qquad
 \lbrace \phi, P_{\phi} \rbrace = 1.
 \label{8}
 \ee
 The canonical momentum for the variable $N(t)$ is
 \be
 P_N \equiv \frac{\partial L}{\partial \dot N} = 0,
 \label{9}
 \ee
 this equation merely constrains the variable $N(t)$.
 The canonical Hamiltonian can be calculated in the usual way, it has
 the form $H_c = NH_0$, then the total Hamiltonian is
 \be
 H_T = NH_0 + u_N P_N,
 \label{10}
 \ee
 where $u_N$ is the Lagrange multiplier associated to the constraint
 $P_N = 0$ in (\ref{9}), and  $H_0$ is the Hamiltonian 
 \be
 H_0 = \left( - \frac{\kappa^2 P^2_R}{12R} + \frac{\pi^2_\phi}{2R^3} -
 \frac{3kR}{\kappa^2} + R^3 V(\phi ) \right).
 \label{11}
 \ee
 The time evolution of any dynamical variables is generated by
 (\ref{10}). For the compatibility of the constraint the Eq. (\ref{9})
 and the dynamics generated by the total Hamiltonian of Eq. (\ref{10}), the
 following equation must hold
 \be
 H_0=0,
 \label{12}
 \ee
 which constrains the dynamics of our system. So, we proceed to the quantum
 mechanics from the above classical system. We introduce the wave function
 of the Universe $\psi$. The constraint equation (\ref{12}) must be
 imposed on the states
 \be
  H_0 \psi = 0.
 \label{13}
 \ee
This constraint nullifies all the dynamical evolution generated by the
total Hamiltonian (\ref{10}). A commutator of any operator and the total
Hamiltonian becomes zero, if it is evaluated for the above constrained
states.The disappearence of time seems disappointing, however, it is a proper
consequence of the invariance of general coordinate transformation in
general relativity. The equation (\ref{9}) merely says, that the wave function
$\psi$ does not depend on the lapse function $N(t)$. Therefore, we expect that
the equation in (\ref{13}) may contain any information of dynamics, since the
WKB solutions of the equation (\ref{13}) is indeed parametrized by an 
``external" time \cite{20}. In the WKB approach the coordinate $T$ is usually 
called WKB time. In quantum
cosmology the constraint (\ref{13}) is known as the Wheeler-DeWitt
equation (time-independent Schr\"odinger equation).

In order to get a time-dependent Schr\"odinger equation we introduce the
time parameter $T(t)$ using the relation
$N(t)= \frac{dT}{dt}$. We shall
regard the following invariant action
\be               
S_r = \frac{1}{\kappa^3} \int R^3 P_T \left( \frac{dT(t)}{dt} -
N(t)\right) dt,
\label{14}
\ee
where $(T, P_T)$ is a pair of canonical variables, $P_T$ is the momentum
 conjugate to $T$. This action is invariant under reparametrization of time
 (\ref{3}), if $P_T$ and $T$ transform as a scalars under reparametrization
 (\ref{3})
 \be
 \delta P_T = a(t) \dot P_T \qquad\qquad \delta T = a(t)\dot T,
 \label{15}
 \ee
 and $N, R$ as in (\ref{4}).

 So, adding the action (\ref{14}) to the action (\ref{6}) we have the total
 invariant action $\tilde S = S_g + S_m + S_r$. In the first order form, we
 get
 \be
 \tilde S = \int \left\{\dot R P_R + \dot \phi P_{\phi} - NH_0
 + \frac{R^3 P_T}{\kappa^3} \left(\frac{dT}{dt} - N(t)\right) \right \} dt.
 \label{16}
 \ee
 We shall proceed with the canonical quantization of the action (\ref{16}).
 We define the canonical momenta $\pi_T$ and $\pi_{P_T}$ corresponding to
the
 variables $T$ and $P_T$, respectively. We get
 \be
 \pi_T \equiv \frac{\partial \tilde L}{\partial \dot T} =
 \frac{R^3}{\kappa^3} P_T, \qquad\qquad
 \pi_{P_T} \equiv \frac{\partial \tilde L}{\partial \dot P_T} = 0,
 \label{17}
 \ee
 leading to the constraints
 \be
 \Pi_1 \equiv \pi_T - \frac{R^3}{\kappa^3} P_T = 0, \qquad\qquad
 \Pi_2 \equiv \pi_{P_T} = 0.
 \label{18}
 \ee

 So, we define the matrix $C_{AB}$, $(A, B =1,2)$ as a Poisson brackets
 between the constraints $C_{AB} = \lbrace \Pi_A, \Pi_B \rbrace$. Then,
 we have the following non-zero matrix elements
 \be
 \lbrace \Pi_{1}, \Pi_{2} \rbrace = -\frac{R^3}{\kappa^3},
 \label{19}
 \ee
 with their inverse matrix elements $(C^{-1})^{1,2} =
-\frac{\kappa^3}{R^3}$.
 The Dirac's brackets $\lbrace , \rbrace^\ast$ are defined by
 \be
 \lbrace f, g \rbrace^{\ast} = \lbrace f, g \rbrace -
 \lbrace f, \sqcap_A \rbrace C^{-1}_{AB} \lbrace \sqcap_B, g \rbrace.
 \label{20}
 \ee
 The result of this procedure leads to the non-zero Dirac's brackets

 \be
 \lbrace T, P_T \rbrace^{\ast} =  \frac{\kappa^3}{R^3}.
 \label{21}
 \ee
 Then, the canonical Hamiltonian is
 \be
 \tilde H_c = N \left(\frac{R^3}{\kappa^3} P_T + H_0 \right ),
 \label{22}
 \ee
 where the Hamiltonian constraint corresponding to the action
 (\ref{16}) is
 \be
 \tilde H =  \frac{ R^3}{\kappa^3} P_T + H_0.
 \label{23}
 \ee

 At the quantum level the Dirac's brackets become commutators
 \be
 [ T, P_T ] = i\lbrace T, P_T \rbrace^{\ast} = i \frac{\kappa^3}{R^3}.
 \label{24}
 \ee
 So, taking the momentum $P_T$ corresponding to $T$ as
 \be
 P_T = -i \frac{\kappa^3}{R^3} \frac{\partial}{\partial T},
 \label{25}
 \ee
 the constraint (\ref{23}) becomes quantum condition on the wave
 function $\psi$,
 \be
 i\frac{\partial}{\partial T} \psi (T,R,\phi )= H \left(- i\frac{\partial}
 {\partial R}, - i\frac{\partial}{\partial\phi},R,\phi \right)\psi.
 \label{26}
 \ee
 There is a question of the factor ordering in the differential equation 
(\ref{26}). In order to find a correct quantum expression for the 
Hamiltonian, we must always consider the factor ordering ambiguites
\be
i\frac{\partial}{\partial T}\psi = \left[ \frac{\kappa^2}{12}
R^{-p-1} \frac{\partial}{\partial R} R^p \frac{\partial}{\partial R}
-\frac{1}{2R^3} \frac{\partial^2}{\partial\varphi^2}
- \frac{3kR}{\kappa^2}+ R^3 V(\varphi )\right] \psi.
 \label{27}
 \ee
 The parameter $p$ takes into account some of the factor ordering ambiguity
 of the theory. The equation (\ref{26}) is the time-dependent Schr\"odinger 
equation for minisuperspace.

 The equations of motion are obtained by demanding that the action
 $\tilde S = S_g + S_m + S_r$ is extremal, $i.e.$ the functional derivatives
 of $\tilde S$ must be zero
 \be
 \frac{\delta \tilde S}{\delta R} = \frac{\delta S_g}{\delta R} +
 \frac{\delta S_m}{\delta R} + \frac{\delta S_r}{\delta R} = 0.
 \label{28}
 \ee
 As a consequence of the equation of motion
 \be
 \frac{\delta \tilde S}{\delta P_T} =
 \frac{\delta S_r}{\delta P_T}= \frac{R^3}{\kappa^3} (\dot T - N) = 0,
 \label{29}
 \ee
 the last term in (\ref{28}) $\frac{3 R^2}{\kappa^3} P_T(\dot T -  N)$
 dissapears and, in fact, inclusion in $S$ of an additional invariant action
 $S_r$ does not change the equations of motion except the equation
 $\frac{\delta \tilde S}{\delta N} = 0$, which is the constraint (\ref{23}).

In the case of general relativity, the canonical quantization makes use of the
$3 + 1$ spliting of the space-time geometry by Arnowit-Deser-Misner (ADM)
\cite{3}. Acording to the ADM prescription of general relativity one consider
 a slicing of the space-time by a family of space-like hypersurfaces labeled
by $t$. This can be thought of as a time coordinate, so that each slice
is identified by the relation $t = const$. So, we introduce a time 
$T(t,x)$, which will be related to time $t$ as
\begin{equation}
n^\mu \partial_\mu T = 1.
\end{equation}
We consider time $T$ as a canonical coordinate. Its corresponding canonical
conjugate momentum will be $P_T$. These $T$ and $P_T$ transform under general
coordinate transformations as scalars. So, in this case, the additional 
action term, which is invariant under general coordinate transformations can
be written in the following form

\begin{eqnarray}
S_{(d=4)} &=& \frac{1}{\kappa^3} \int \sqrt{-g} P_T(n^\mu \partial_\mu T -1)
d^4 x\nonumber\\
&=& \frac{1}{\kappa^3} \int Nh^{1/2} P_T \Big(\frac{\partial_0 T}{N} -
\frac{N^i\partial_i T}{N} - 1 \Big) dt d^3x \label{31}\\
&=& \frac{1}{\kappa^3} \int h^{1/2} P_T (\partial_0 T - N^i \partial_i T -
N )dt d^3x, \nonumber
\end{eqnarray}
varying this action with respect to the three metric $h_{ik}$ and $P_T$, we get

\begin{eqnarray}
\frac{\delta S_{(d=4)}}{\delta h_{ik}} &=&  \frac{1}{2} 
\frac{h^{1/2}}{\kappa^3}
h^{ik} P_T (\partial_0 T - N^i \partial_i T - N ) = 0 \nonumber \\
\frac{\delta S_{(d=4)}}{\delta P_T} &=&  \frac{h^{1/2}}{\kappa^3} 
(\partial_0 T - N^i \partial_i T - N) = 0.
\end{eqnarray}
So, given a four dimensional space-time geometry with the metric $g_{\mu \nu}$
considered as a parameter family of three-dimensional space-like hypersurfaces
$t =const$ the intrinsic metric of each surface is  $h_{i j} = g_{i j}$,
$g = \det{ g_{\mu \nu}}$, $h = \det{h_{i j}}$ and $g = Nh$. The unit 
future-directed normal vector is $n^\mu$, $( n^\mu n_\mu = -1)$ to
hypersurface $t=const$ with component $n_\mu = (-N, 0, 0, 0)$ and
$n^\mu = (\frac{1}{N}, -\frac{N^i}{N})$, where  $N^i$ is the shift vector,
(for the metric (\ref{1}) the shift vector is $N^i = 0$) and
$h^{1/2} = R^3$.

So, if we consider the four dimensional gravity interacting with a scalar
matter field and the invariant additional term (\ref{31}) after applying the
ADM spliting $(3+1)$ formalism for the FRW model, we get

\begin{eqnarray}
S &=& - \frac{1}{2 \kappa^2} \int \sqrt{-g} R d^4 x - 
\int \sqrt{-g}\Big[ \frac{(\partial_\mu \phi)^2}{2} + V(\phi) \Big] d^4 x
\nonumber\\
&+& \frac{1}{\kappa^3}\int \sqrt{-g} P_T (n^\mu \partial_\mu T - 1) d^4 x =
\int \Big[ \Big( -\frac{3R {\dot R}^2}{2N \kappa^2} + \frac{3}{\kappa^2}
kNR \Big) \label{32}\\
&+& \frac{R^3 {\dot \phi}^2}{2N} - NR^3 V(\phi) \Big] dt + 
\frac{1}{\kappa^3} \int R^3 P_T 
\Big( \frac{dT(t)}{dt} - N(t) \Big) dt.\nonumber 
\end{eqnarray}
In particular, putting the gauge $N=1$, then $T=t$ and we obtain the so-called
cosmic time, on the other hand, if we take $N=\frac{R}{\kappa}$ then we get
the conformal time gauge. In terms of the $(3 + 1)$ variables, the action 
(\ref{32}) takes the form \cite{20}
\begin{eqnarray}
S &=& \frac{1}{2\kappa^2}\int N h^{1/2}\Big(K_{ij} K^{ij} - K^2 + 
^{(3)}R\Big) dt d^3x \label{32a}\\
&& + \frac{1}{\kappa^3} \int h^{1/2} P_T \Big(\partial_0 T - N^i 
\partial_i T - N \Big) dt d^3x + S_{matter},\nonumber     
\end{eqnarray}
where $K$ is the trace of the extrinsic curvature $K_{ij}$. In the action
(\ref{32a}) we have ignored the surface term. Then, the canonical Hamiltonian 
is

\begin{eqnarray}
{\tilde H} &=& N\Big(-\frac{h^{1/2}}{\kappa^3} P_T + 
\frac{\kappa^2}{2} G_{ijkl} \pi^{ij} \pi^{kl} - \frac{1}{2\kappa^2} 
(h^{1/2}) {^{(3)}R} + H_{matter}(\phi, \pi_{\phi}) \Big)\label{32b}\\
&& + N^i \Big( \frac{1}{\kappa^3} P_T \partial_i T - 2 D_j \pi_i^j +
H_{i (matter)}\Big), \nonumber
\end{eqnarray}
where $\pi^{ij}$ and $\pi_{\phi}$ are the momenta conjugated to $h_{ij}$ and
$\phi$, respectively. $D_i$ is a covariant derivative on the metric $h_{ij}$
and $G_{ijkl} = \frac{1}{2} h^{-1/2}(h_{ik} h_{jl} + h_{il}h_{jk} - 
h_{ij}h_{kl})$. The Dirac quantization of this model will lead to the
many-fingered time Schr\"odinger equation (Tomanaga-Shwinger equation) for
the wave function $\Psi(T, h_{ik}, \phi)$

\begin{eqnarray}
i\frac{\delta \Psi}{\delta T} &=& \Big( -\frac{1}{2\kappa} 
G_{ijkl} \frac{\delta}{\delta h_{ij}} \frac{\delta}{\delta{h_{kl}}} + 
\frac{1}{2\kappa} b_{ik} \frac{\delta}{\delta h_{ik}} 
- \frac{1}{2\kappa^3} h^{-1/2} \frac{\delta^2}{\delta \phi^2} \label{32c}\\
&& - \frac{\kappa}{2}(h^{1/2}) {^{(3)}R} + \frac{\kappa^3}{2}h^{1/2}h^{ik} 
\phi_{,i} \phi_{,k} + \kappa^3 V(\phi) \Big), \nonumber
\end{eqnarray}
where the coefficients $b_{ik}$ depend of the chose of factor ordering in 
the term $-\frac{1}{2\kappa} G_{ijkl} \frac{\delta}{\delta h_{ij}} 
\frac{\delta}{\delta h_{kl}}$. 

\section{Supersymmetric FRW Model}
 In order to obtain a superfield formulation of the action (\ref{6})
 the transformation of the time reparametrization (\ref{3}) must
 be extended to the $n=2$ local conformal time supersymmetry (LCTS)
$(t, \eta, \bar\eta)$ \cite{21,22}. These transformations can be
 written as
 \begin{eqnarray}
 \delta t &=& {I\!\!L}(t, \theta,\bar\theta) + \frac{1}{2}\bar\theta
 D_{\bar\theta} {I\!\!L}(t,\theta,\bar\theta) - \frac{1}{2}\theta
D_{\theta}
 {I\!\!L}(t,\theta,\bar\theta), \nonumber\\
 \delta \theta &=& \frac{i}{2} D_{\theta} {I\!\!L}(t,\theta,\bar\theta),
 \qquad \delta \bar\theta = \frac{i}{2} D_{\bar\theta}
 {I\!\!L}(t,\theta,\bar\theta),
 \label{33}
 \end{eqnarray}
 with the superfunction ${I\!\!L}(t,\theta,\bar\theta)$ defined by
 \be
 {I\!\!L}(t,\theta,\bar\theta) = a(t) + i\theta {\bar\beta}^\prime(t) +
 i\bar\theta \beta^\prime(t) + b(t)\theta \bar\theta,
 \label{34}
 \ee
 where $D_{\theta} = \frac{\partial}{\partial \theta} + i\bar\theta
 \frac{\partial}{\partial t}$ and $D_{\bar\theta} = - \frac{\partial}
 {\partial \bar\theta} - i\theta \frac{\partial}{\partial t}$ are the
 supercovariant derivatives of the $n=2$ supersymmetry, $a(t)$ is a local
 time reparametrization parameter, $\beta^\prime(t)$ is the Grassmann
 complex parameter of the local conformal $n=2$ supersymmetry
transformations
 and $b(t)$ is the parameter of the local $U(1)$ rotations on the Grassmann
 coordinates $\theta$ $(\bar\theta = \theta^\dagger)$.
 Then, the superfield generalization of the action (\ref{6}), which is
 invariant under the $n=2$ (LCTS) (\ref{33}) has the form
 \cite{23,24}
 \begin{eqnarray}
 S_{(n=2)} &=& S_g + S_m = \int \left( -\frac{3}{\kappa^2}{I\!\!N}^{-1}
 {I\!\!R} D_{\bar\theta} {I\!\!R} D_{\theta}{I\!\!R} +
 \frac{3 \sqrt k}{\kappa^2} {I\!\!R}^2 \right)
 d\theta d\bar\theta dt \label{35}\\
 &+& \int \left (\frac{1}{2}{I\!\!N}^{-1} {I\!\!R}^3 D_{\bar\theta}
 {\bf \Phi} D_{\theta} {\bf \Phi} - 2 {I\!\!R}^3 g(\bf \Phi) \right )
 d\theta d\bar\theta dt, \nonumber
 \end{eqnarray}
 where $g(\Phi)$ is the superpotential. The most general supersymmetric
 interaction for a set of complex homogeneous scalar fields with the
 scale factor was considered in \cite{25,26}. For the one-dimensional gravity superfield ${I\!\!N}(t,\theta,\bar\theta)$
 we have the following series expansion
 \be
 {I\!\!N}(t,\theta,\bar\theta) = N(t) + i\theta {\bar\psi}^\prime(t) +
 i\bar\theta \psi^\prime(t) + V^\prime(t) \theta\bar\theta,
 \label{36}
 \ee
 where $N(t)$ is the lapse function, $\psi^\prime = N^{1/2} \psi$
 and $V^\prime = N V + \bar\psi \psi$. The components
 $N, \psi, \bar\psi$ and $V$ in (\ref{36}) are gauge fields of the
 one-dimensional $n=2$ supergravity. The superfield (\ref{36}) transforms
as one-dimensional vector under the (LCTS) (\ref{33}),
 \be
 \delta {I\!\!N} = ({I\!\!L}{I\!\!N})^. + \frac{i}{2}D_{\bar\theta}
 {I\!\!L}D_{\theta} {I\!\!N} + \frac{i}{2}D_{\theta}{I\!\!L} D_{\bar\theta}
 {I\!\!N}.
 \label{37}
 \ee
 The series expansion for the superfield ${I\!\!R}(t,\theta,\bar\theta)$
has the form
 \be
 {I\!\!R}(t,\theta,\bar\theta) = R(t) + i\theta {\bar\lambda}^\prime(t) +
 i\bar\theta \lambda^\prime(t) + B^\prime(t) \theta\bar\theta,
 \label{38}
 \ee
 where $R(t)$ is the scale factor of the FRW Universe,
 $\lambda^\prime = \kappa N^{1/2} \lambda$ and
 $B^\prime= \kappa N B + \frac{\kappa}{2}(\bar\psi \lambda -
 \psi \bar\lambda)$.

 For the real scalar matter superfield ${\bf \Phi}(t,\theta,\bar\theta)$ we
 have
 \be
 {\bf \Phi}(t,\theta,\bar\theta) = \phi(t) + i\theta \bar\chi^\prime(t) +
 i\bar\theta \chi^\prime(t) + F^\prime(t) \theta\bar\theta,
 \label{39}
 \ee
 where $\chi^\prime = N^{1/2}\chi$ and
 $F^\prime = N F + \frac{1}{2}(\bar\psi \bar\chi -
 \psi \chi)$. The components $B(t)$ and $F(t)$ in the superfields
 ${I\!\!R}$ and ${\bf \Phi}$ are auxiliary fields. The superfields
(\ref{38})
 and (\ref{39}) transform as scalars under the transformations (\ref{33}).

 Performing the integration over $\theta, \bar\theta$ in (\ref{35}) and
 eliminating the auxiliary fields $B$ and $F$ by means of their equations
of motion, the action (\ref{35}) takes its component form. The first-class
 constraints may be obtained from the component form of the action
 (\ref{35}) varying it with respect to $N(t), \psi(t), \bar\psi(t)$ and
 $V(t)$, respectively. Then, we obtain the following first-class
constraints
 $H_0=0$, $S=0$, $\bar S= 0$ and $F=0$, where
 \begin{eqnarray}
 H_0 &=& - \frac{\kappa^2}{12} \frac{\pi^2_R}{R} - \frac{3kR}{\kappa^2}
 -\frac{\sqrt k}{3R} \bar\lambda \lambda + \frac{\pi^2_\phi}{2R^3} -
 \frac{i\kappa}{2R^3} \pi_\phi (\bar\lambda \chi + \lambda \bar\chi)
 - \frac{\kappa^2}{4R^3} \bar\lambda \lambda \bar\chi \chi \nonumber\\
 &+& \frac{3 \sqrt k}{2R} \bar\chi \chi + \kappa^2 g(\phi) \bar\lambda
\lambda
 + 6 \sqrt k g(\phi)R^2 + 2 \left(\frac{\partial g}{\partial \phi}\right)^2
 R^3 - 3 \kappa^2 g^2(\phi) R^3 \nonumber\\
 &+& \frac{3}{2} \kappa^2 g(\phi) \bar\chi\chi
 + 2 \frac{\partial^2 g}{\partial \phi^2} \bar\chi \chi +
 \kappa \frac{\partial g}{\partial \phi}(\bar\lambda \chi - \lambda
\bar\chi),
 \label{40}
 \end{eqnarray}
 \begin{eqnarray}
 S&=& \left(\frac{i\kappa}{3} R^{-1/2} \pi_R - \frac{2\sqrt k }
 {\kappa}R^{1/2} + 2\kappa g(\phi) R^{3/2} + \frac{\kappa}{4} R^{-3/2}
 \bar\chi \chi \right) \lambda \nonumber\\
 &+& \left( iR^{-3/2} \pi_\phi + 2 R^{3/2}
 \frac{\partial g}{\partial \phi}\right ) \chi,
 \label{41}
 \end{eqnarray}
 and
 \be
 F = - \frac{2}{3}\bar\lambda \lambda + \bar\chi \chi,
 \label{42}
 \ee
 where $\bar S = S^\dagger$.

 The canonical Hamiltonian is the sum of all the constraints
 \be
 H_{c(n=2)} = NH_0 + \frac{1}{2} \bar\psi S - \frac{1}{2}\psi \bar S +
 \frac{1}{2} V F.
 \label{43}
 \ee
 In terms of Dirac's brackets for the canonical variables $R, \pi_R, \phi,
 \pi_\phi, \lambda, \bar\lambda, \chi$ and $\bar\chi$ the quantities $H_0,
 S,\bar S$ and $F$ form the closed super-algebra of conserving charges
 \begin{eqnarray}
 \lbrace S, \bar S \rbrace^\ast &=& - 2iH_0, \qquad
 \lbrack H_0, S \rbrack^\ast = \lbrack H_0, \bar S \rbrack^\ast = 0
 \label{44} \\
 \lbrack F, S\rbrack^\ast &=& iS, \qquad
 \lbrack F, \bar S \rbrack^\ast = -i\bar S. \nonumber
 \end{eqnarray}
 So, any physically allowed states must obey the following quantum
constraints
 \begin{eqnarray}
 H_0 \psi &=& 0, \qquad S\psi = 0, \qquad \bar S \psi = 0, \qquad
 F \psi = 0,
 \label{45}
 \end{eqnarray}
 when we change the classical variables by their corresponding operators.
 The first equation in (\ref{45}) is the Wheeler-DeWitt equation for the
 minisuperspace model. Therefore, we have the time-independent
Schr\"odinger
 equation, this fact is due to the invariance under reparametrization
 symmetry of the action (\ref{35}), this problem is well-known as the
 ``problem of time" \cite{1} in the minisuperspace models and general
 relativity theory. Due to the super-algebra (\ref{44}) the second and
 the thirth equations in (\ref{45}) reflect the
 fact, that there is a ``square root" of the Hamiltonian $H_0$ with zero
 energy states. The constraints Hamiltonian $H_0$, supercharges
 $S, \bar S$, $F$ follow from the invariance of the action (\ref{35})
 under the $n=2$ (LCTS) transformations (\ref{33}).

 In order to have a time-dependent Schr\"odinger equation for the
 supersymmetric minisuperspace models with the action (\ref{35})
 we consider a generalization of the reparametrization invariant action
 $S_r$ (\ref{14}). In the case of $n=2$ (LCTS) it has the superfield form
 \begin{eqnarray}
 S_{r(n=2)} &=& -\int \left[ {I\!\!P} -\frac{i}{2} {I\!\!N}^{-1}
 \left( D_{\bar\theta} {\bf T} D_\theta{I\!\!P} - D_{\bar\theta}
 {I\!\!P}D_\theta {\bf T} \right)\right] d\theta d\bar\theta dt.
 \label{46}
 \end{eqnarray}
 Note, that the $Ber E^A_B$, as well as the superjacobian of $n=2$
 (LCTS) transformations, is equal to one and is omitted in the actions
 (\ref{35},\ref{46}). The action (\ref{46}) is determined in terms of the
new
 superfields ${\bf T}$ and ${I\!\!P}$. The series expansion for ${\bf T}$
 has the form
 \be
 {\bf T}(t,\theta,\bar\theta) = T(t) + \theta \eta^\prime(t) -
 \bar\theta \bar\eta^\prime(t) + m^\prime(t) \theta\bar\theta,
 \label{47}
 \ee
 where $ \eta^\prime = N^{1/2} \eta$ and $m^\prime =
 N m + \frac{i}{2}(\bar\psi \bar\eta + \psi \eta)$.
 The superfield ${\bf T}$ is determined by the odd complex time variables
 $\eta(t)$ and $\bar\eta(t)$, which are the superpartners of the time
 $T(t)$ and one auxiliary parameter $m(t)$.

 The transformation rule for the superfield ${\bf T}(t,\theta,\bar\theta)$
 under the $n=2$ (LCTS) transformations is
 \be
 \delta {\bf T} = {I\!\!L} \dot{\bf T} + \frac{i}{2}D_{\bar\theta}
 {I\!\!L} D_\theta {\bf T} + \frac{i}{2}D_\theta {I\!\!L}D_{\bar\theta}
 {\bf T},
 \label{48}
 \ee
 and transforms as a scalar under the transformations (\ref{33}). The
 superfield ${I\!\!P}(t,\theta,\bar\theta)$ has the form
 \be
 {I\!\!P}(t,\theta,\bar\theta) = \rho(t) + i\theta P^\prime_{\bar\eta}(t)
 + i\bar\theta P^\prime_\eta(t) + P_T(t) \theta\bar\theta,
 \label{49}
 \ee
 where $P^\prime_\eta = N^{1/2} P_\eta$ and $P^\prime_T =
 N P_T + \frac{1}{2}(\bar\psi P_{\eta} -
 \psi P_{\bar\eta})$, $P_\eta$ and $P_{\bar\eta}$ are the odd
 complex momenta, $i.e.$ the superpartners of the momentum $P_T$.

 The superfield ${I\!\!P}(t,\theta,\bar\theta)$ transforms as
 \be
 \delta {I\!\!P}(t,\theta,\bar\theta) = {I\!\!L} \dot {I\!\!P} +
 \frac{i}{2}D_{\bar\theta} {I\!\!L} D_\theta {I\!\!P} +
 \frac{i}{2} D_\theta {I\!\!L} D_{\bar\theta} {I\!\!P}.
 \label{50}
 \ee
 The action (\ref{46}) is invariant under the $n=2$ (LCTS) transformtions
 (\ref{33}). Performing the integration over $\theta$ and $\bar\theta$ in
 (\ref{46}) and making the redefinitions $P_T \to \frac{R^3}{\kappa^3}P_T$,
 $P_\eta \to \frac{R^3}{\kappa^3}P_\eta$ and
 $P_{\bar\eta} \to \frac{R^3}{\kappa^3}P_{\bar\eta}$ we obtain the
component action
 \begin{eqnarray}
 S_{r(n=2)} &=& -\int \left\{ \frac{R^3}{\kappa^3} \left(P_T(N - \dot T) +
 i\dot \eta P_\eta + i\dot{\bar\eta} P_{\bar\eta} +
 \frac{\bar\psi}{2}(P_\eta - \bar\eta P_T) \right. \right.\label{51}\\
 &-& \left. \left. \frac{\psi}{2}(P_{\bar\eta} - \eta P_T) +
 \frac{V}{2}(\eta P_\eta - \bar\eta P_{\bar\eta})\right) + m\dot \rho
 -\frac{i}{2}m\psi P_{\bar\eta} - \frac{i}{2} m \bar\psi P_\eta \right\}dt.
 \nonumber
 \end{eqnarray}
 We can see from (\ref{51}) that the momenta $P_{\eta}, P_{\bar\eta}$
 and $P_T$ in the superfield (\ref{49}) are related with the components
 of the superfield (\ref{36}), which enter in the action (\ref{35}),
 unlike those momenta, the component $\rho$ of the superfield (\ref{49})
 and the component of the superfield (\ref{47}) are not related with
 any components in (\ref{36}). Hence, one can show that the variables
 $\rho$ and $m$ are auxiliary fields in the sense, that they can be
 eliminated from (\ref{51}) by some unitary transformation after this, we
 have 
 \begin{eqnarray}
 S_{r(n=2)} &=& - \int \frac{R^3}{\kappa^3}\left\{ P_T(N - \dot T) +
 i\dot \eta P_\eta + i\dot{\bar \eta}P_{\bar\eta} + \frac{\bar\psi}{2}
 (P_\eta - \bar\eta P_T) \right. \nonumber\\
 &-& \left. \frac{\psi}{2}(P_{\bar\eta} - \eta P_T) +
 \frac{V}{2}(\eta P_\eta - \bar\eta P_{\bar\eta}) \right\} dt.
 \label{52}
 \end{eqnarray}
 In addition to the canonical momenta $\pi_T$ and $\pi_{P_T}$
 for the two even constraints (\ref{17}), the action (\ref{52}) has
 the additional momenta ${\cal P}_\eta$ and ${\cal P}_{P_\eta}$
 conjugate to $\eta$ and $P_\eta$, respectively,
 \begin{eqnarray}
 {\cal P}_\eta &=& \frac{\partial L_{r(n=2)}}{\partial\dot\eta} =
 -i \frac{R^3}{\kappa^3} P_\eta, \qquad
 {\cal P}_{P_\eta} = \frac{\partial L_{r(n=2)}}{\partial \dot P_\eta} = 0.
 \label{53}
 \end{eqnarray}
 With respect to the canonical odd Poisson brackets we have
 \be
 \lbrace \eta , {\cal P}_\eta \rbrace = 1, \qquad
 \lbrace P_\eta , {\cal P}_{P_\eta} \rbrace = 1.
 \label{54}
 \ee
 They form two primary constraints of the second class
 \be
 \sqcap_3 (\eta ) \equiv {\cal P}_\eta +i\frac{R^3}{\kappa^3} P_\eta = 0 ,
 \qquad \sqcap_4 (P_{\eta})\equiv {\cal P}_{P_{\eta}} = 0.
 \label{55}
 \ee
 The only non-vanishing Poisson bracket between these constraints is
 \be
 \{ \sqcap_3 , \sqcap_4 \} = i \frac{R^3}{\kappa^3} .
 \label{56}
 \ee
 The momenta ${\cal P}_{\bar\eta}$ and ${\cal P}_{P_{\bar\eta}}$ conjugate
to
 $\bar\eta$ and $P_{\bar\eta}$ respectively, it also gives two primary
 constraints of the second-class
 \be
 \sqcap_5 (\bar\eta) \equiv {\cal P}_{\bar\eta}+i\frac{R^3}{\kappa^3}
 P_{\bar\eta} = 0 ,\qquad \sqcap_6 (P_{\bar\eta}) =
 {\cal P}_{P_{\bar\eta}} = 0,
 \label{57}
 \ee
 with non-vanishing Poisson bracket
 \be
 \{\sqcap_5, \sqcap_6 \} = i \frac{R^3}{\kappa^3}.
 \label{58}
 \ee
 The constraints (\ref{55}) and (\ref{57}) for the Grassmann dynamical
 variables can be eliminated by Dirac's procedure. Defining the matrix
 constraint $C_{ik} (i,k =\eta ,P_\eta ,\bar\eta ,P_{\bar\eta})$ as the
 odd Poisson bracket we have the following non-zero matrix elements
 \begin{eqnarray}
 C_{\eta P_\eta} &=& C_{P_\eta \eta} =\{ \sqcap_3,
 \sqcap_4\}=i\frac{R^3}{\kappa^3}, \nonumber\\
 C_{{\bar\eta} P_{\bar\eta}} &=& C_{P_{\bar\eta\bar\eta}}
 =\{\sqcap_5,\sqcap_6 \} = i \frac{R^3}{\kappa^3},
 \label{59}
 \end{eqnarray}
 with their inverse matrices $(C^{-1})^{\eta P_\eta}=-i
\frac{\kappa^3}{R^3}$
 and $(C^{-1})^{\bar\eta P_\eta}=-i \frac{\kappa^3}{R^3}$.
 The result of this procedure is the elimination of the momenta conjugate
to
 the Grassmann variables, leaving us with the following non-zero Dirac's
 bracket relations
 \be
 \lbrace \eta ,P_\eta \rbrace^\ast = i\frac{\kappa^3}{R^3}, \qquad
 \lbrace \bar\eta , P_{\bar\eta}\rbrace^\ast = i\frac{\kappa^3}{R^3}.
 \label{60}
 \ee
 So, if we take the additional term (\ref{46}), then the full action is
 \be
 \tilde S_{(n=2)} = S_{(n=2)} + S_{r(n=2)}.
 \label{61}
 \ee
 The canonical Hamiltonian for the action (\ref{61}) will have the
 following form
 \begin{eqnarray}
 \tilde H_{c(n=2)} &=& N\left (\frac{R^3}{\kappa^3} P_T + H_0 \right)
 + \frac{\bar\psi}{2} \left (-\frac{R^3}{\kappa^3} S_\eta + S
\right)\nonumber\\
 &-& \frac{\psi}{2} \left (\frac{R^3}{\kappa^3} S_{\bar\eta} + \bar S
\right)
 + \frac{V}{2} \left ( \frac{R^3}{\kappa^3} F_\eta + F \right),
 \label{62}
 \end{eqnarray}
 where $S_\eta = (-P_\eta + \eta P_T)$, $S_{\bar\eta} = (P_{\bar\eta} -
 \eta P_T)$, $F_\eta = (\eta P_\eta - \bar\eta P_{\bar\eta})$, and
 $H_0, S, \bar S$ and $F$ are defined in (\ref{40},\ref{41},\ref{42}). In
 the component form of the action (\ref{61}) there are no kinetic terms
 for $N, \psi, \bar\psi$ and $V$. This fact is reflected in the primary
 constraints $P_N =0$, $P_\psi = 0$, $P_{\bar\psi} = 0$ and $P_V = 0$,
 where $P_N, P_\psi, P_{\bar\psi}$ and $P_V$ are the canonical momenta
 conjugated to $N, \psi, \bar\psi$ and $V$, respectively. Then, the total
 Hamiltonian may be written as
 \be
 \tilde H = \tilde H_{c(n=2)} + u_NP_N + u_\psi P_\psi +
 u_{\bar\psi} P_{\bar\psi} + u_V P_V.
 \label{63}
 \ee
 Due to the conditions $\dot P_N = \dot P_\psi = \dot P_{\bar\psi} =
 \dot P_V = 0$ we now have the first-class constraints 
 \begin{eqnarray}
 \tilde H &=& \frac{R^3}{\kappa^3}P_T + H_0 = 0, \qquad
 {\cal F} = \frac{R^3}{\kappa^3} F_\eta + F = 0, \nonumber\\
 Q_\eta &=& -\frac{R^3}{\kappa^3} S_\eta + S = 0, \qquad
 Q_{\bar\eta} =  \frac{R^3}{\kappa^3} S_{\bar\eta} + \bar S = 0.
 \label{64}
 \end{eqnarray}
 They form a closed super-algebra with respect to the Dirac's brackets
 \begin{eqnarray}
 \lbrace Q_\eta, Q_{\bar\eta} \rbrace^\ast &=& - 2i \tilde H, \qquad
 \lbrack \tilde H, Q_\eta \rbrack^\ast =
 \lbrack \tilde H, Q_{\bar\eta} \rbrack^\ast = 0 \nonumber\\
 \lbrack {\cal F}, Q_\eta \rbrack^\ast &=& iQ_\eta, \qquad
 \lbrack {\cal F}, Q_{\bar\eta} \rbrack^\ast = -i Q_{\bar\eta}.
 \label{65}
 \end{eqnarray}
 After quantization Dirac's brackets must be replaced by
 anticommutators
 \be
 \lbrace \eta , P_\eta \rbrace = i\lbrace \eta, P_\eta \rbrace^\ast =
 -\frac{\kappa^3}{R^3},\qquad
 \lbrace \bar\eta, P_{\bar\eta}\rbrace =
 i \lbrace \bar\eta, P_{\bar\eta} \rbrace^\ast = -\frac{\kappa^3}{R^3},
 \label{66}
 \ee
 with the operator relations
 \begin{eqnarray}
 P_\eta &=& -\frac{\kappa^3}{R^3} \frac{\partial}{\partial\eta}, \qquad
 P_{\bar\eta} = -\frac{\kappa^3}{R^3} \frac{\partial}{\partial\bar\eta} .
 \label{67}
 \end{eqnarray}
 To obtain the quantum expression for $H_0, S, \bar S, F$ we
 must solve the operator ordering ambiguity. Such ambiguities always take
 place when the operator expression contains the product of non-commuting
 operators $\lambda$ and $\bar\lambda,\chi$ and $\bar\chi$ $R$ and
 $\pi_{_{\bar R}} = - i\frac{\partial}{\partial R}, \phi$ and $\pi_\phi
= -i \frac{\partial}{\partial\phi}$, such procedure leads in our case to the
 following expressions for the generators on the quantum level
 \begin{eqnarray}
 \tilde H &=&-i\frac{\partial}{\partial T} + H_0 (R,\pi_R,\phi , \pi_\phi
 , \lambda , \bar\lambda ,\chi \bar\chi) , \nonumber \\
 Q_\eta &=&
 -\Big(\frac{\partial}{\partial\eta}-i\bar\eta\frac{\partial}{\partial T}
 \Big)+S(R,\pi_R ,\phi ,\pi_\phi ,\lambda ,\chi) , \label{68} \\
 Q_{\bar\eta} &=&  \Big(-\frac{\partial}{\partial\bar\eta} +i\eta
 \frac{\partial}{\partial T} \Big)+\bar S (R, \pi_R ,\phi ,\pi_\phi
 ,\bar\lambda ,\bar\chi ) ,\nonumber \\
 {\cal F} &=& \Big(-\eta \frac{\partial}{\partial\eta}+\bar\eta
 \frac{\partial}{\partial\bar\eta} \Big) + F(\lambda ,\bar\lambda ,\chi
 ,\bar\chi ), \nonumber
 \end{eqnarray}
 where $S_\eta =\frac{\partial}{\partial\eta} - i\bar\eta
 \frac{\partial}{\partial T}$ and $S_{\bar\eta} = -
 \frac{\partial}{\partial\bar\eta} +i\eta
 \frac{\partial}{\partial T}$ are the generators of the supertranslation,
 $ P_T =-i \frac{\partial}{\partial T}$ is the ordinary time translation
 on the superspace with coordinates $(t,\eta,\bar\eta)$
 \be
 \{ S_\eta , S_{\bar\eta} \} =2i \frac{\partial}{\partial T} ,
 \label{69}
 \ee
 and $F_\eta = -\eta \frac{\partial}{\partial\eta} + \bar\eta
 \frac{\partial}{\partial\bar\eta}$ is the ${\cal U}$(1) generator of the
rotation
 on the complex Grassmann coordinate $\eta (\bar\eta = \eta^\dagger)$.
 The algebra of the quantum generators of the conserving charges $H_0 ,S,
 \bar S, F$ is a closed super-algebra
 \begin{eqnarray}
 \lbrace S,\bar S\rbrace &=&2H_0,\quad \lbrack S, H_0\rbrack =
 \lbrack S,H_0\rbrack = \lbrack \tilde S,H_0 \rbrack =
 \lbrack F,H_0\rbrack=0
 \nonumber \\
 S^2 =\bar S^2 &=&0,\qquad [ F,S]=- \bar S,\qquad [F, \bar S] = \bar S.
 \label{70}
 \end{eqnarray}
 We can see from Eqs. (\ref{65}) and (\ref{68}) that the operators
 $\tilde H, Q_\eta , Q_{\bar\eta}$ and ${\cal F}$ obey the same
super-algebra
 (\ref{70})
 \begin{eqnarray}
 \lbrace Q_\eta ,Q_{\bar\eta}\rbrace &=& 2\tilde H,\quad
 \lbrack Q_\eta ,\tilde H \rbrack = \lbrack Q_\eta ,\tilde H \rbrack =
 \lbrack {\cal F},\tilde H\rbrack= 0
 \nonumber \\
 Q^2_\eta &=& Q^2_{\bar\eta} = 0,\qquad \lbrack {\cal F}, Q_\eta \rbrack =
 - Q_\eta ,\quad \lbrack{\cal F} , Q_{\bar\eta}\rbrack = Q_{\bar\eta} .
 \label{71}
 \end{eqnarray}
 In the quantum theory the first-class constraints (\ref{68}) become
 conditions on the wave function $\Psi$, which has the superfield form
 \begin{eqnarray}
 \Psi (T,\eta,\bar\eta, R, \phi, \bar\phi,\lambda,\bar\lambda,
 \chi, \bar\chi ) &=& \psi (T, R, \phi,\lambda,\bar\lambda,\chi,\bar\chi )
 \nonumber\\
 &+& i\eta \xi (T,R,\phi ,\lambda ,\bar\lambda ,\chi ,\bar\chi )
 + i\bar\eta \zeta (T,R,\phi , \lambda ,\bar\lambda , \chi ,\bar\chi )
 \nonumber\\
 &+& \sigma (T,R,\phi , \lambda , \bar\lambda , \chi , \bar\chi )
 \eta\bar\eta.
 \label{72}
 \end{eqnarray}

 So, we have the supersymmetric quantum constraints
 \be
 \tilde H \Psi = 0, \qquad Q_\eta \Psi = 0, \qquad Q_{\bar\eta} \Psi = 0,
 \qquad {\cal F} \Psi = 0.
 \label{73}
 \ee
 Taking the constraints
 \be
 Q_\eta \Psi = 0, \qquad Q_{\bar\eta} \Psi = 0,
 \label{74}
 \ee
 and due to the algebra (\ref{71})
 \be
 \lbrace Q_\eta, Q_{\bar\eta} \rbrace \Psi = 2 \tilde H \Psi = 0.
 \label{75}
 \ee
 This is a time-dependent Schr\"odinger equation for the
 minisuperspace model.

 The condition (\ref{75}) leads to the following form for the wave function
 (\ref{72})
 \be
 \psi_\ast = \psi + \eta ( S \psi) + \bar\eta (\bar S \psi)-
 \frac{1}{2} (\bar SS -S\bar S) \psi \eta\bar\eta ,
 \label{76}
 \ee
 then $Q_\eta \psi_*$ has the following form
 \begin{eqnarray}
 Q_\eta \psi_* &=& \bar\eta \Big(i\frac{d\psi}{dT} -\frac{1}{2} \{S,
 \bar S\} \psi \Big)+ \nonumber \\
 &+& \eta\bar\eta S\Big(i\frac{d \psi}{dT}-\frac{1}{2} \{S,\bar S\}
 \psi\Big) =0,
 \label{77}
 \end{eqnarray}
 this is the standard Schr\"odinger equation and due to the relation
 $H_0 =\frac{1}{2} \{ S,\bar S\}$ it may be written as
 \be
 i\frac{\partial \psi}{\partial T} = H_0 \psi,
 \label{78}
 \ee
 where the wave function is
 $\psi (T, R, \lambda ,\bar\lambda ,\chi ,\bar\chi )$. If we put in the
 Schr\"odinger equation (\ref{78}) the condition of a stationary state
 given by $\frac{\partial \psi}{\partial T} = 0$, we will have that
 $H_0 \psi =0$ and due to the algebra (\ref{70}) we obtain $S \psi = \bar
 S\psi = 0$ and the wave function $\psi_\ast$ becomes $\psi$.

 \vspace{.5cm}
 \noindent {\bf Acknowledgments.} We are grateful to E. Ivanov, S.
Krivonos, J.L. Lucio, I. Lyanzuridi, L. Marsheva, O. Obreg\'on, 
M.P. Ryan, J. Socorro and M. Tsulaia for their interest in the work and 
useful comments. This research was supported in part by CONACyT under the 
grant 28454E. Work of A.I.P. was supported in part by INTAS grant 96-0538 and 
by the Russian Foundation of Basic Research, grant 99-02-18417. One of use 
J.J.R. would like to thank CONACyT for support under Estancias Posdoctorales 
en el Extranjero and Instituto de F\'{\i}sica de la Universidad de Gto. for its
 hospitality during the final stages of this work.


 \end{document}